\documentclass{article}
\usepackage{ijcai22}

\usepackage{times}

\usepackage{soul}
\usepackage{url}
\usepackage[hidelinks]{hyperref}
\usepackage[utf8]{inputenc}
\usepackage[small]{caption}
\usepackage{graphicx}
\usepackage{amsmath}
\usepackage{booktabs}
\usepackage{amssymb}
\usepackage{subfig}
\usepackage{mathtools}
\usepackage[margin=3cm]{geometry}
\usepackage[linesnumbered,lined,boxed]{algorithm2e}
\usepackage[english]{babel}
\usepackage{amsthm}

\theoremstyle{definition}
\newtheorem{definition}{Definition}[section]

\DeclareMathOperator*{\argmax}{arg\,max}

\title{Quantifying Health Inequalities Induced by Data and AI Models}
\author{
Honghan Wu$^1$\footnote{Contact Author}
\and
Minhong Wang$^1$\and
Aneeta Sylolypavan$^{1}$\And
Sarah Wild$^2$
\\
\affiliations
$^1$Institute of Health Informatics, University College London, London, United Kingdom
\\
$^2$Usher Institute, University of Edinburgh, Edinburgh, United Kingdom
\emails
\{honghan.wu, minhong.wang, aneeta.sylolypavan\}@ucl.ac.uk, sarah.wild@ed.ac.uk
}

\begin{document}

\maketitle

\begin{abstract}
AI technologies are being increasingly tested and applied in critical environments including healthcare. Without an effective way to detect and mitigate AI induced inequalities, AI might do more harm than good, potentially leading to the widening of underlying inequalities.
This paper proposes a generic allocation-deterioration framework for detecting and quantifying AI induced inequality. Specifically, AI induced inequalities are quantified as the area between two allocation-deterioration curves. 
To assess the framework's performance, experiments were conducted on ten synthetic datasets (N$>$33,000) generated from HiRID - a real-world Intensive Care Unit (ICU) dataset, showing its ability to accurately detect and quantify inequality proportionally to controlled inequalities. Extensive analyses were carried out to quantify health inequalities (a) embedded in two real-world ICU datasets; (b) induced by AI models trained for two resource allocation scenarios. Results showed that compared to men, women had up to $33\%$ poorer deterioration in markers of prognosis when admitted to HiRID ICUs. All four AI models assessed were shown to induce significant inequalities ($2.45\%$ to $43.2\%$) for non-White compared to White patients. The models exacerbated data embedded inequalities significantly in 3 out of 8 assessments, one of which was $>$9 times worse.

The codebase is at \url{https://github.com/knowlab/DAindex-Framework}.
\end{abstract}

\section{Introduction}

Artificial intelligence (AI) technologies in medicine hold great potential to facilitate better decision-making and efficient service delivery in health and social care, hence, widely and highly expected to improve clinical outcomes in the near future~\cite{Topol2019-mw}. However, a critical and alarming caveat is that AI driven decision making systems, particularly those using data-driven technologies, are subject to, or themselves cause, bias and discrimination that may exacerbate existing health inequity among racial and ethnicity groups~\cite{leslie2021does}. 

“Bias in, bias out” is the catchphrase used to highlight concerns about the fact that data driven AI models make inferences by finding ‘patterns’ from the data they analyse. As racial and ethnic disparities have long existed in health and care~\cite{Nelson2002-xv,Van_Ryn2000-zk}, inferences from such biased data would inevitably channel embedded inequality into decisions or suggestions they derive. So, effective mitigation is required~\cite{Bailey2017-tq}. 

Training data might induce bias even when there is no embedded inequalities from service deliveries. Under-representation of minority groups in datasets creates a real technical challenge for data driven approaches to draw sensible conclusions for such groups, creating another probable cause of inequality exacerbated by AI. Small samples of a minority group will cause computational models to draw inaccurate predictions for them~\cite{Rajkomar2018-nz}. 

In addition to biases rooted in the data, further bias could arise in the whole pipeline of AI development and deployment. In particular, health inequalities might be introduced through model selection, the feature engineering process (the choice of the input variables) and label determinations (the choice of target variables)~\cite{Passi2019-ba}. 

To mitigate these inequalities, conceptual frameworks have been proposed~\cite{Rajkomar2018-nz}, and qualitative analysis and checklist based guidance have been suggested~\cite{Vyas2020-mj}. While these tools are useful for understanding the possible types of biases and where they might arise, these solutions are not able to \emph{quantify health inequalities} so that AI practitioners can debug, evaluate and audit potential biases in data, model developments and deployments.

All in all, we currently lack effective technical tools for AI practitioners to deal with the critical issue of detecting and mitigating  health inequalities embedded in the training data and/or induced by the AI models' developments and deployments. In this paper, 
\begin{itemize}
    \item we propose a generic quantification framework called \emph{allocation-deterioration indices} that can quantify health inequalities from both datasets and AI models.  It is available at \url{https://github.com/knowlab/DAindex-Framework}.
    \item we propose novel and pragmatic solutions to address several technical issues associated with the deterioration index computation including boundary bias of kernel density estimation and challenges associated with discrete random variables.
    \item ten synthetic datasets (N$>$33,000) and two real-world health datasets (total N$>$70,000) are used and three clinical decision-making scenarios are tested in an extensive set of experiments.
\end{itemize}

\section{Method}
%The lack of a generic and standardised quantifiable framework, equivalent to metrics like the Receiver Operating Characteristic (ROC) Curve for analysing medical diagnostic tests~\cite{Hajian-Tilaki2013-xf}, is one of the key technical barriers in addressing bias in AI for medicine. Without it, the definition and evaluation of health inequalities need to be done by individual studies. This impedes the efficient and compatible detection, evaluation, comparison and communication of health inequalities in datasets and AI models among the health data science community.

\begin{figure}[t]
\includegraphics[width=6.5cm]{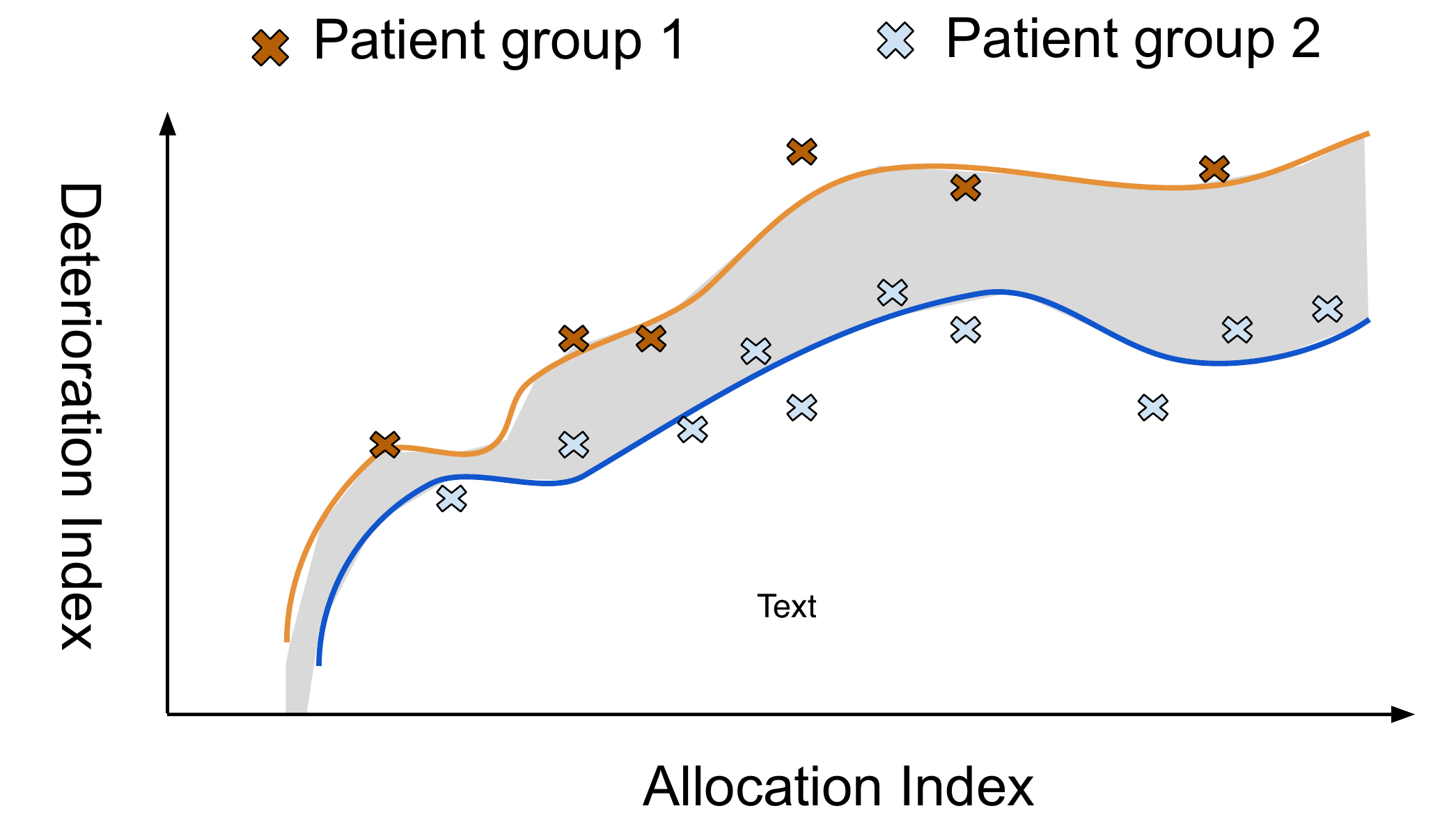}
\caption{Area between Allocation-Deterioration curves: a generic inequality quantification metric for AI models and derivation data}
\label{fig:DAIndex}
\end{figure}

Inspired by~\cite{Obermeyer2019-pq} and generalising from it, we define and quantify health inequalities in a generic resource allocation scenario using a so-called allocation-deterioration framework, as conceptualised in Figure~\ref{fig:DAIndex}. 

The basic idea is to define two indices: allocation index and deterioration index. The allocation index is (to be derived) from the AI model of interest. Conceptually, AI models are abstracted as ``resource allocators", such as predicting the probability of Intensive Care Unit admission. Note that the models themselves do not need to be particularly designed to allocate resources, for example, it could be risk prediction of cardiovascular disease (CVD) among people with diabetes~\cite{Dinh2019-oi}. Essentially, a resource allocator is a computational model that takes patient data as input and outputs a (normalised) score between 0 and 1. We call this score the allocation index. Deterioration index is a score between 0 and 1 to measure the deterioration status of patients. It can be derived from an objective measurement for disease prognosis (i.e., \textbf{a marker of prognosis} in epidemiology terminology), such as extensively used comorbidity scores~\cite{De_Groot2003-uv,Obermeyer2019-pq} or biomarker measurements like those for CVDs~\cite{Vasan2006-ag}. 

When we have the two indices, each patient can then be represented as a point in a two-dimensional space of $(\textit{allocation index}, \textit{deterioration index})$, as illustrated in Figure~\ref{fig:DAIndex}. A group of patients are then translated into a set of points in the space, for which a regression model could be fitted to approximate as a curve in the space. The same could be done for another group. \emph{The area between the two curves is then the deterioration difference between their corresponding patient groups, quantifying the inequalities induced by the ``allocator", i.e., the AI model that produces the allocation index}. The curve with the larger area under it represents the patient group which would be unfairly treated if the allocation index was to be used in allocating resources or services: a patient from this group would be deemed healthier than a patient from another group who is equally ill. The rest of this section gives technical details of realising key components of this conceptual framework.

\subsection{Deterioration Index Definition}
%How to quantify a deterioration index with a value from 0 to 1? How to deal with different types of measurements: discrete and continuous? How to quantify binary cut-offs and severity-aware measurements? 

For a group of patients $P=\{p_1, p_2, ..., p_n\}$, a deterioration index is a function $d \colon \mathbb{P}(P) \xrightarrow[]{m} [0, 1]$, where the value $1$ denotes the most deteriorated. $m$ stands for a numeric measurement function $m\colon P \rightarrow \mathbb{R}$, where $p \in P$. For example, it can be counting the number of multimorbidities of a patient or the heart beat rates. Generally, let $\{m(p)|p \in P \}$ be $M=\{M_1, M_2,...,M_n\}$, an independent and identically distributed sample.

The deterioration status is usually quantified as the degree to which the measured value is in excess of what is normal. For example, the normal range of Creatinine (a measure for kidney functions) for adult men is 0.74 to 1.35 mg/dL~\footnote{\url{https://www.mayoclinic.org/tests-procedures/creatinine-test/about/pac-20384646}}. A reading of 5 mg/dL is apparently off-the-scale, however, it is probably quantified as less deteriorated than a reading of 10 mg/dL. Following this idea, without loss of generality, we quantify the deterioration index as $$d(P;m)=f(\{M_1, M_2,...,M_n\};t_m)$$ where $f$ quantifies the degree to which these measures are in excess of a threshold $t_m$.

A simple implementation, as defined in Definition~\ref{def:1cutoff}, is to use the probability of $M$ having a value greater than a given cut-off $t_m$. This is intuitive as it quantifies the likelihood of having abnormal measurements within a patient group. For example, when using Creatinine as the measurement, $t_m$ can be set as $1.35$, the upper bound of normal readings for men. A group of patients with $f_{Pr}=0.6$ is more deteriorated than another group with $f_{Pr}=0.3$ in terms of their kidney functions.

\begin{definition}[Probability beyond one cut-off]
\label{def:1cutoff}
Let \(f_{Pr}\) be an implementation of $f$, as $Pr(M \geq t_m)$ where $Pr$ stands for a probability function.
\end{definition}

However, $f_{Pr}$ is not able to discriminate a group with a distribution more skewed to the far end of the spectrum from another with a distribution closer to the cut-off when both having the same $f_{Pr}$. For example, for two groups with Creatinine measures as $\{0.8, 0.78, 10\}$ and $\{0.8, 0.78, 1.36\}$ respectively, both will have the same $f_{Pr}(M;1.35)=0.3$, while the former is clearly more deteriorated as it has a much higher abnormal reading. 

To address this issue, we propose a `probability beyond $k$-step cut-offs' as defined in Definition~\ref{def:kstep}. It splits the relevant value range 
%($[t_m, max_m]$, $max_m$ is the maximum possible value of measurement $m$), 
into $k$ steps and allows the quantification to put more weights on more deteriorated values using a weight function.

\begin{definition}[Probability beyond $k$-step cut-offs]
\label{def:kstep}
Let $k$ a constant integer and \(f_{Pr}^{k}\) be an implementation of $f$, as defined below 
$$
\small \sum_{i=1}^{k}{w(i) \cdot Pr\big((t_m + (i-1) \cdot \delta) \leq M < (t_m + i \cdot \delta)\big)}
$$ where $\delta=\lceil \frac{max_m-t_m}{k} \rceil$, $max_m$ is the maximum possible value of $m$ and $w(i) \rightarrow \mathbb{R}$ is a weight function which meets $\sum_{i=1}^{k}{w(i)}=1$.
\end{definition}

Let $t_m=1.35$, $k=2$ and $w(1)=0.3; w(2)=0.7$, the above two groups will have $f_{Pr}^{2}$ values of $0.21$ and $0.09$, respectively.
\subsection{Deterioration Index Implementation}
\begin{figure*}[t]
\subfloat{\includegraphics[width=7.6cm]{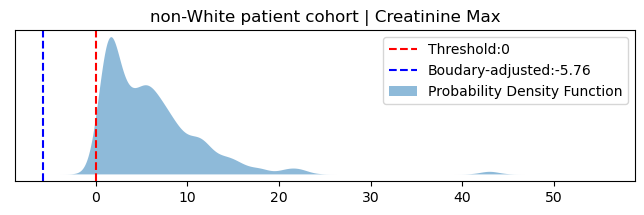} }%
% \qquad
\subfloat{\includegraphics[width=7.6cm]{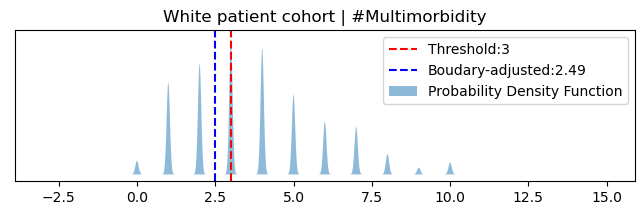} }%
\caption{Kernel Density Estimation Issues. The left figure illustrates the boundary bias. It plots a PDF estimated for maximum Creatinine readings (ranged from 0 to 50) of a patient cohort from the MIMIC-III dataset. There is a nonzero probability mass region to the left of legitimate minimal value of zero. The right figure illustrates the issue with pulse-like PDFs for discrete random variables. The PDF was estimated from the number of multimorbitidies of a white patient cohort. Clearly, each discrete value has two regions with nonzero probability mass to either direction.}
\label{fig:dis-boundary}
\end{figure*}
%Kernel Density Estimation, the quantification (treatment on pulse shaped PDF, weighted sum on integrated severity), boundary bias issue
To obtain the probabilities in the above definitions, we adopt a kernel density estimation approach, which is a standard non-parametric method for estimating a probability density. Let $g_M$ be the probability density function (PDF) of a measurement random sample $M$. A kernel density estimator (KDE) is $\hat{g}_M(v;h)=\frac{1}{nh}\sum_{i=1}^{n}K(\frac{v-M_i}{h})$, where $K$ is the kernel function and $h$ is the bandwidth parameter used for smoothing the estimate. We use a Gaussian kernel $K(v)=exp(-v^2/2)/\sqrt{2 \pi}$ in our implementation, as Gaussianity has been assessed over diverse settings and has shown good performances in our experiments.

However, there is a well-known boundary bias issue~\cite{geenens2014probit} with the kernel density estimation. That is when the random variable is bounded to a closed interval, the KDE will exhibit significant bias at at the end-points of the interval because $\hat{g}_M(v;h)$ will have nonzero probability mass outside the interval. Unfortunately, in the clinical domain, almost all measurements are bounded to a closed interval. The figure at the left of Figure~\ref{fig:dis-boundary} illustrates one example of such a case. It is the PDF plot for the \emph{Creatinine Max} distribution of a non-White adult patient cohort from the MIMIC-III dataset~\cite{johnson2016mimic}. The PDF is estimated using a Gaussian KDE with a bandwidth of $1.0$, which was chosen from a hyper-parameter tuning with grid search. The range of the values is $[0, 50]$. Clearly, there is a nonzero region to the left of the minimal possible value 0 (the red dashed vertical line). Using such a PDF to calculate $Pr(M \geq 0)$ would certainly lead to an inaccurate probability. 

In addition to boundary issues at intervals, a much-less studied issue but fairly prevalent in health domain is the pulse-like PDFs which are often estimated for discrete random variables, such as number of multimorbidities of patients. Technically, they are associated with small bandwidth values learned for KDE models. The right figure in Figure~\ref{fig:dis-boundary} is such an example, which is a PDF estimated for numbers of multimorbidities for a White patient group from MIMIC-III. The grid-searched optimal bandwidth for this random sample was $0.0526$. It presents a pulse-like PDF with peaks around possible discrete values. To calculate above defined deterioration indices (e.g., $f_{Pr}^k$) using $t_m=3$, one would need to get the probability $Pr(M \geq 3)$. Using the PDF as it is would lead to an inaccurate result because the nonzero probability mass right to the left of the cutoff (the red dotted line in the figure) is relevant but would be ignored. 

While the boundary issue associated with closed intervals has been studied in the literature for more than several decades, existing approaches are either not very generalisable or difficult to implement~\cite{colbrook2020kernel}. In fact, very few have been implemented in R or Python libraries. Moreover, few of them tackle the issue associated with pulse-like PDFs as described above. To address these issues, we propose a pragmatic, automated boundary adjustment approach. Algorithm ~\ref{alg:adjb} 
in the appendix describes the adjustment for the left boundaries including pulse-like PDFs. This is needed for accurately estimating $Pr(M\geq t_m)$. The similar logic could be applied for the right boundary adjustment for $Pr(M\leq t_m)$. The blue dashed lines in Figure~\ref{fig:dis-boundary} denote the adjusted values for given thresholds.

% \RestyleAlgo{ruled}
% \SetKwComment{Comment}{/* }{ */}

% \begin{algorithm}[hbt!]
% \caption{Left Boundary Adjustment}\label{alg:adjb}
% \SetKwData{Left}{left}\SetKwData{This}{this}\SetKwData{Up}{up}
% \SetKwFunction{Union}{Union}\SetKwFunction{FindCompress}{FindCompress}
% \SetKwInOut{Input}{input}\SetKwInOut{Output}{output}

% \Input {$E$: learned KDE\; $lb$: the lower bound\; 
% $ub$: the upper bound\; 
% $t$: value to adjust\; 
% $t_p$: $\argmax(\{v|v \in M: v < t\})$ when $M$ is discrete and $t$ is not boundary, otherwise  $t$\; 
% $\varepsilon$: a small constant like $1^{-10}$\; 
% $V$: an empty array.} 
% \Output {$\hat{t}$: the adjusted value for $t$}
% \If{$len(V) = 0$}{
% \Comment{get an evenly spaced numbers between $lb$ and $ub$ with a relatively big number $n$, e.g., $n=20 \times (ub - lb)$.}
% $a \gets gen(lb, ub, n)$\;
% $s \gets (ub - lb) / n$\;
% \For{$i\leftarrow 1$ \KwTo len($a$)}{
% $x_p \gets lb$\;
% \If{$i > 1$}{
% $x_p \gets a[i-1]$\;
% }
% $x \gets a[i]$\;
% $p \gets exp(E(x))$\;
% \While{$p \geq \varepsilon$ and $x > x_p$}{
% $x \gets (x - s)$\;
% $p \gets exp(E(x))$\;
% }
% \If{$exp(E(x)) < \varepsilon$}{
% $V.add(x)$\;
% }
% }
% }
% $\hat{t} \gets \argmax(\{v|v \in V\colon v < t\})$\;
% \If{$\hat{t} \leq t_p$}{
% $\hat{t} \gets t$\;
% }
% return $\hat{t}$\; 
% \end{algorithm}

\subsection{Area under allocation-deterioration curve}
The first step to get the area is to generate the allocation-deterioration curve (A-D curve for short). To do that, we start with a \emph{resource allocator}, which in this context, as you will recall, is essentially an AI model used for decision making. Technically, a \emph{resource allocator} is $a(p) \in [0, 1]$ assigning a score for quantifying the degree of a patient $p$ needs some service/resource. 

\begin{definition}[Allocation-Deterioration Curve]
\label{def:adcurve}
Given a measurement $m$, an allocator $a$ and a deterioration index $d$, the allocation-deterioration curve is defined as 
$$
\forall x \in [0, 1], \big (x, d(\{p|p\in P, a(X(p))=x\};m)\big).
$$

\end{definition}

In reality, for a particular dataset, the set of patients having one particular allocation score $x$ might be empty or too few to obtain a reliable estimation of their deterioration status. To address this, we propose an approximation method as described in Algorithm~\ref{alg:curve} in the appendix. %~\ref{alg:curve}.

% \begin{algorithm}[]
% \caption{Approximate A-D Curve}\label{alg:curve}
% \SetKwInOut{Input}{input}\SetKwInOut{Output}{output}
% \Input {$P$: the patient cohort\;
% $d$: the deterioration index function\;
% $m$: the measurement\;
% $a$: the AI model as an allocator\;
% $l$: a constant for smoothing\; 
% $n$: a constant for specifying the number of points to be generated\;
% $\nu$: a threshold for the minimal numbers of patients for deterioration estimation. 
% } 
% \Output {the curve}
% $C \gets []$\;
% $X \gets gen(0, 1, n)$\;
% \For{$x$ in $X$}{
% $\hat{P} \gets \{p|p\in P: (x-l) \leq a(p) < (x+l)\}$\;
% \If{$|\hat{P}|\geq \nu$}{
% $C.add((x, d(\hat{P};m)))$\;
% }
% }
% return $C$
% \end{algorithm}

For those missed points in our approximation, interpolation techniques~\cite{muschelli2020roc} could be applied to fill the blank. However, the missing data does not affect inequality quantification as it is calculated as the relative difference between two groups of patients in the same dataset. Keeping the missingness reflects the actual characteristics of the cohorts, leading to an accurate result.

The area under the A-D curve can be estimated using numerical integration, using simple geometric shapes to approximate the area under the curve. We choose Simpson's rule~\footnote{\url{https://personal.math.ubc.ca/~pwalls/math-python/integration/simpsons-rule/}}.
\subsection{Inequality quantification}
Finally, we define the inequality between two groups of patients. Definition~\ref{def:dbineq} defines the inequality embedded in a dataset. 
\begin{definition}[Inequality embedded in a dataset]
\label{def:dbineq}
Given two patient groups $P_1$ and $P_2$ being assigned a resource, a measurement $m$, and a deterioration index function $d(P;m)$, the inequality of $P_1$ compared to $P_2$ (denoted as \textbf{$P_1$ vs $P_2$}) is quantified as $\frac{d(P_1;m)}{d(P_2;m)} - 1$.
\end{definition}

%Let's define $AUC$ as: $$AUC = \int\frac{d(\{p|p\in P, a(p)=x\};m)}{|\{p|p\in P, a(p)=x\}|} \,dx.$$ 
For AI induced inequality, let $AUC(a, P, d, m)$ be the area under the A-D curve of model $a$ for $P$ using $d(P;m)$ as deterioration index. Let $AUC(a, P, d, m;\tau)$ be the area of the sub-region where the allocation index $\geq \tau$. The AI induced inequality can then be defined in Definition~\ref{def:aiineq}.
\begin{definition}[Inequality induced by a model]
\label{def:aiineq}
In a decision making scenario with an allocation threshold $\tau$, given a model $a$, patient groups $P_1$ and $P_2$, a measurement $m$, and a deterioration index function $d(P;m)$, the inequality of $P_1$ over $P_2$ induced by $a$ is quantified as $$
\frac{AUC(a, P_1, d, m;\tau)}{AUC(a, P_2, d, m;\tau)} - 1 .
$$
\end{definition}
\section{Results}
\subsection{Datasets and Cohorts}
Two real-world intensive care unit (ICU) datasets were used for experiments, namely: (1) HiRID: a freely accessible critical care dataset containing de-identified data for $>$33,000 ICU admissions to the Bern University Hospital, Switzerland, between 2008-2016~\cite{faltys2021hirid}; (2) MIMIC-III: a freely available database containing de-identified data for $>$40,000 ICU patients of the Beth Israel Deaconess Medical Centre, Boston, United States, between 2001-2012~\cite{johnson2016mimic}. Refer to the appendix for ethic statements.
% \begin{itemize}
%     \item HiRID (v1.1.1): a freely accessible critical care dataset containing de-identified data for $>33,000$ ICU admissions to the Bern University Hospital, Switzerland, between 2008-2016~\cite{faltys2021hirid}.  
%     \item MIMIC-III (v1.4): a freely available database containing de-identified data for $>40,000$ ICU patients of the Beth Israel Deaconess Medical Centre, Boston, United States, between 2001-2012~\cite{johnson2016mimic}.
% \end{itemize}

Two case-control cohorts were extracted from MIMIC-III, each of which was for analysing a resource allocation scenario of deciding the need for surgery: (1) Renal Autotransplantation: 146 patients were identified using the ICD-9-CM Procedure Code $55.69$. A control cohort (N=438) was then matched up using 1:3 ratio based on ethnicity, gender and age (+/- 3 years). The total cohort size is 584; (2) Operations on Kidney: 584 patients were identified using the ICD-9-CM Procedure Code 55.xx, where `x' means wildcard. A similar control matching method was used and identified 1,752 control patients. The total cohort size is 2,336.
% \begin{itemize}
%  \item Renal Autotransplantation: 146 patients were identified using the ICD-9-CM Procedure Code $55.69$. A control cohort (N=438) was then matched up using $1:3$ ratio based on ethnicity, gender and age (+/- 3 years). The total cohort size is 584.
%  \item Operations on Kidney: 584 patients were identified using the ICD-9-CM Procedure Code $55.xx$, where $x$ means wildcard. A similar control matching method was used and identified 1,752 control patients. The total cohort size is 2,336.
% \end{itemize}

% \begin{enumerate}
%     \item data embedded
%     \item under-representation: (a) class imbalance; (b) minor groups' characteristics.
%     \item model development: feature selections; target variables
% \end{enumerate}

% \usepackage{multirow}
% \usepackage{multirow}

\subsection{Inequality Quantification Evaluation}
% \begin{figure*}[]
% \caption{Inequality Quantification Evaluation on synthetic data: y-axis is the inequality quantity of female vs male. x-axis is the percentage of controlled improvements on readings of the female subcohort.  Y-value of each point is the mean value of 10 runs on the same x-value, i.e., \% of improvement. Shaded areas denote 25-75\% quantile regions.}
% \label{fig:increaseEval}
% \begin{center}
% \includegraphics[width=12cm]{figs/inequality_evaluation.png}
% \end{center}
% \end{figure*}
We conducted a few experiments to check whether and how our inequality model works. Specifically, we wanted to evaluate: (a) when there was no bias or inequality, would our model correctly detect it? (b) could our model accurately quantify the known percentages of inequalities? To mimic a near real-world situation, we used the HiRID dataset to generate synthetic data. The generation process was composed of (1) randomly select 10\% data from HiRID and choose all male patients out of it; (2) randomly change the sex of 50\% of the patients to female. 

We evaluated the inequality associated with ICU admissions. Specifically, we used Definition~\ref{def:dbineq} to assess inequality of \textbf{female vs male} at a \emph{resource allocation scenario} of \textbf{ICU admission}. Three measurements (prognosis markers) were chosen for quantifying deterioration indices: \textit{Creatinine max value}, \textit{Creatinine min value} and \textit{ALT min value}. We selected readings with the first 24 hours of admission. \textit{Creatinine} measures kidney functions and normal ranges chosen were: 65.4 to 119.3 micromoles/L for women and 52.2 to 91.9 micromoles/L for men. \textit{ALT} measures liver functions and normal ranges chosen were: $\leq 30$ U/L for men and $\leq 19$ U/L for women~\cite{kunde2005spectrum}. The deterioration index used a \emph{probability on 20-step cut-offs}.

For answering the above question (a), i.e., detecting no inequality, we generated 10 synthetic datasets using the above-mentioned process and ran inequality assessments on these datasets. Note that the synthetic data were actual data of male patients. Therefore, with sufficient numbers of sampling, there should NOT be any significant amount of inequality between male and (synthetically created) female patients overall. Appendix Table~\ref{tab:10synth} shows the overall results of 10 runs on 10 such datasets. The $p$-value was generated for a T-test for the null hypothesis that the mean value was equal to 0, meaning NO inequality. We observed all $p$-values were not significant, meaning we could not reject the null hypothesis. This means in all three measurements, the mean values of 10 runs were equal to 0, indicating our model quantified no significant inequalities.

For the above question (b), i.e., whether our model could quantify inequality proportionally to actual inequality, we used the same process as the previous experiment of generating 10 datasets. Then, for female patients, we purposely improved their measurements by changing readings towards the more healthier end, e.g., decrease the Creatinine max readings, increase Creatinine min readings. We selected different levels of improvements - 10 steps evenly spaced between 0.0 and 0.5. For each of them, we quantified the inequality of female vs male. Figure~\ref{fig:increaseEval} in the Appendix depicts the results of this experiment. In all cases, the model correctly identified the level of inequality changes - inequality trends going downwards consistently when the strength of improvement increases. Specifically, the Spearman rank-order correlation coefficients between the inequality quantities and the percentages of improvements are -0.989, -0.974 and -0.993 for Creatinine Max/Min and ALT Max respectively, showing near perfect negative correlations. 

% \begin{table}[]
% \begin{tabular}{ccl}
% \multicolumn{3}{c}{Health inequality assessments on synthetic datasets}                                                                                      \\ \hline
% \multicolumn{1}{l}{Measurement}                                                & mean [95\% CI]                                  & $p$-value \\ \hline \hline
% \multicolumn{1}{c|}{\begin{tabular}[c]{@{}c@{}}Creatinine \\ max\end{tabular}} & \multicolumn{1}{c|}{0.044 {[}-0.083,  0.130{]}} & 0.0664  \\ \hline
% \multicolumn{1}{c|}{\begin{tabular}[c]{@{}c@{}}Creatinine \\ min\end{tabular}} & \multicolumn{1}{c|}{0.024 {[}-0.266,  0.302{]}} & 0.7084  \\ \hline
% \multicolumn{1}{c|}{\begin{tabular}[c]{@{}c@{}}ALT \\ max\end{tabular}}        & \multicolumn{1}{c|}{0.033 {[}-0.157,  0.182{]}} & 0.4231  \\ \hline
% \end{tabular}
% \caption{Overall inequality of \textbf{female vs male} quantified on 10 synthetic datasets, where there should be no inequality overall.}
% \label{tab:10synth}
% \end{table}

\subsection{Dataset embedded inequality analysis}
% \begin{figure}[]
% % \includegraphics[width=8.5cm]{discrete-sample-boundary.png}
% \subfloat{\includegraphics[width=7.71cm]{figs/fig_kindney_op_case_nw_female_creatinine.png} }%
% \qquad
% \subfloat{\includegraphics[width=7.71cm]{figs/fig_kindney_op_case_w_female_creatinine.png} }%
% % \qquad
% % \subfloat{\includegraphics[width=7.71cm]{figs/fig_kindney_op_case_nw_male_creatinine.png} }%
% % % \qquad
% % \subfloat{\includegraphics[width=7.71cm]{figs/fig_kindney_op_case_w_male_creatinine.png} }%
% \caption{Probability density functions for quantifying inequalities of \textbf{non-White vs White} in the scenario of kidney operations in MIMIC-III dataset. Dashed lines denote thresholds (i.e., boundary values of abnormal readings) for computing deterioration index. Shaded area are regions where the probability integral happens for getting the deterioration index. The above two figures are females, which illustrate an inequality of 35.06\%. The bottom two are males, where there is an inequality of 19.94\%.}
% \label{fig:mimiccreatinine}
% \end{figure}
The first analysis was conducted on a HiRID dataset for detecting and quantifying its embedded inequality, using the same inequality quantification setup as the previous subsection. 

Table~\ref{tab:hiridcohorts} shows the results on 10 randomly selected cohorts from the total HiRID admissions. Each cohort had 3,390 patients. Among the three deterioration indices, females were shown slightly healthier on \emph{Creatinine max}, but overall marginal with a $p$-value of 0.02. On the other two indices (both with a much higher statistical significance), females were clearly worse off in those two measurements. In particular, they were significantly more ill than male patients on \emph{Creatinine min}, quantified as 0.337 (intepretable as 33.7\% more deteriorated than males at admission). Overall, females admitted to ICU were more deteriorated compared to males within HiRID.

\begin{table}[]
\begin{tabular}{ccl}
\multicolumn{3}{c}{Health Inequality embedded in HiRID dataset}                                                                                      \\ \hline
\multicolumn{1}{l}{Measurement}                                                & mean [95\% CI]                                  & $p$-value \\ \hline \hline
\multicolumn{1}{c|}{\begin{tabular}[c]{@{}c@{}}Creatinine \\ max\end{tabular}} & \multicolumn{1}{c|}{-0.079 {[}-0.207, 0.034{]}} & 0.0219  \\ \hline
\multicolumn{1}{c|}{\begin{tabular}[c]{@{}c@{}}Creatinine \\ min\end{tabular}} & \multicolumn{1}{c|}{0.337 {[}0.181, 0.472{]}} & 0.0000  \\ \hline
\multicolumn{1}{c|}{\begin{tabular}[c]{@{}c@{}}ALT \\ max\end{tabular}}        & \multicolumn{1}{c|}{0.093 {[}0.018, 0.197{]}} & 0.0012  \\ \hline
\end{tabular}
\caption{Inequality analysis of \textbf{Female vs Male} on ten sub-cohorts randomly sampled from HiRID, each with 10\% (N=3,390) of the total patients. The resource allocation scenario is ICU admission and three deterioration indices adopt \textit{probability beyond 20-step cut-offs}, using measurements of \textit{Creatinine max/min} and \textit{ALT max}, respectively.}
\label{tab:hiridcohorts}
\end{table}

The second analysis was conducted on the MIMIC-III dataset to evaluate the health inequality of \textbf{non-White patients vs White patients} at a resource allocation scenario of \textbf{Operations on Kidney} - a cohort with 2,336 patients as described above. We were interested in finding the inequality among the 584 patients who underwent kidney operations. Here, we report the \emph{Creatinine Max} based health inequality. Male and female have different normal ranges (MIMIC III uses mg/dL as the unit):  0.74 to 1.35 for men and 0.59 to 1.04 for women. Therefore, we compare male and female separately. 
%as shown in Figure~\ref{fig:mimiccreatinine} 
Figure~\ref{fig:mimiccreatinine} in the appendix
depicts the PDF distributions of four sub-cohorts. This experiment also used a deterioration index based on the probability beyond 20-step cut-offs. For those who underwent kidney operations in MIMIC-III, female none-White had a 35.06\% inequality over their White female peers, while none-White males were also worse off compared to White males, quantified as 19.94\%. This indicates non-White patients were consistently and substantially more deteriorated in terms of their kidney functions in such a resource allocation scenario. In particular, among all the four subgroups, \textbf{the inequality of non-White male vs White female} was the most significant: 46.57\%.

\subsection{Model induced inequality analysis}

\begin{table}[]
\begin{tabular}{rcccc}
\hline
\multicolumn{5}{c}{Kidney operation}                                                                                                                                                            \\ \hline
\multicolumn{1}{l}{Measurement} & \multicolumn{2}{c|}{\begin{tabular}[c]{@{}c@{}}Creatinine \\ Max\end{tabular}} & \multicolumn{2}{c}{\begin{tabular}[c]{@{}c@{}}Normalised\\ MM\end{tabular}} \\ \hline
DB Inequality                    & \multicolumn{2}{c|}{29.10\%}                                                   & \multicolumn{2}{c}{7.62\%}                                                  \\ \hline
Models                           & LR                                         & \multicolumn{1}{c|}{RF}           & LR                                        & RF                              \\ \hline
DR Inequality                    & \multicolumn{1}{r}{\textbf{37.6\%}}        & \multicolumn{1}{r|}{22.2\%}       & \multicolumn{1}{r}{\textbf{10.52\%}}       & \multicolumn{1}{r}{4.54\%}       \\ \hline
\end{tabular}
\caption{Inequality of \textbf{non-White vs White patients} induced by AI models for predicting kidney operations. \emph{DB inequality} row gives the DB embedded inequality quantities. \emph{DR Inequality} is quantified by areas under A-D curves in the region where a model suggests surgery.}
\label{tab:modelieq1}
\end{table}

\begin{figure}[h]
\center
% \includegraphics[width=8.5cm]{discrete-sample-boundary.png}
% \subfloat{
\includegraphics[width=6cm]{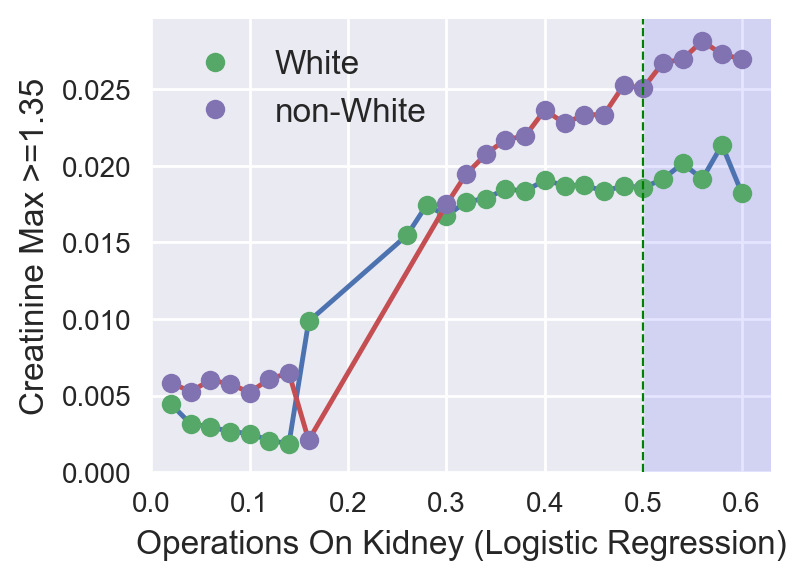} 
% }%
% \subfloat{\includegraphics[width=6cm]{figs/model/fig_model_kidney_op_rf_creatinine_1.35.png} }%
% % \subfloat{\includegraphics[width=3.8cm]{figs/model/fig_model_kidney_op_lr_NMM_3.png} }%
% % \subfloat{\includegraphics[width=3.8cm]{figs/model/fig_model_kidney_op_rf_NMM_3.png} }%
% \qquad
% \subfloat{\includegraphics[width=3.8cm]{figs/model/fig_model_trans_lr_creatinine_1.35.png} }%
% \subfloat{\includegraphics[width=3.8cm]{figs/model/fig_model_trans_rf_creatinine_1.35.png} }%
% \subfloat{\includegraphics[width=3.8cm]{figs/model/fig_model_trans_lr_NMM_3.png} }%
% \subfloat{\includegraphics[width=3.8cm]{figs/model/fig_model_trans_rf_NMM_3.png} }%
\caption{Allocation-Deterioration Curves (\textbf{non-White vs White}) of a model trained for predicting the need for kidney related surgeries. Non-White patients are significantly ($37.6\%$) more severe  within the decision region (shaded area, allocation index $> 0.5$).}
\label{fig:modelineq1}
\end{figure}

To assess the impact of AI models on health inequality if they were used for clinical decision making, two case studies were conducted in two resource allocation scenarios: one on \textbf{Renal Autotransplantation} and the other on more general \textbf{Operations on Kidney}. Tables describing these patient characteristics are available in the appendix (Table~\ref{tab:ptch1} and~\ref{tab:ptch2}).

Logistic Regression (LR) and Random Forest (RF) models were developed for predicting the need for surgeries in both cases with 10-fold cross validation and grid search for hyper-parameter tuning. The two algorithms were chosen because they were widely used in clinical studies. Details of feature selection and hyper-parameters are available in the appendix (Table~\ref{tab:hyperp}). For prediction performances (ROCAUC), LR achieved 0.795 (IQR:0.784-0.805) and 0.867 (IQR:0.843-0.891) for \textit{Operations on Kidney} and \textit{Renal Autotransplantation}, respectively, while RF achieved 0.830 (0.816-0.844) and 0.878 (0.853-0.904), respectively.

For quantifying the inequality, two deterioration indices were used including \emph{Creatinine Max} and a new measurement of \emph{Noramlised number of multimorbidities}, \emph{Normalised MM} for short. The multimorbidities included those in a list of 17 chronic conditions as defined by~\cite{st2021implementing}). \emph{Normalised MM} is defined as $\#MM \times \frac{65}{age}$, where $\#MM$ is the number of multimorbidities a patient had.

\begin{figure*}[h]
\subfloat{\includegraphics[width=3.8cm]{figs/model/fig_model_kidney_op_lr_creatinine_1_35.png} }%
\subfloat{\includegraphics[width=3.8cm]{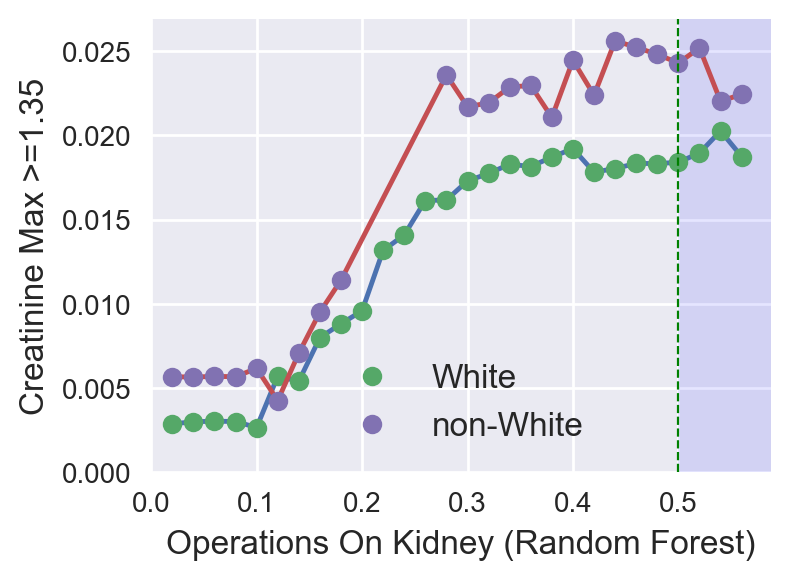} }%
\subfloat{\includegraphics[width=3.8cm]{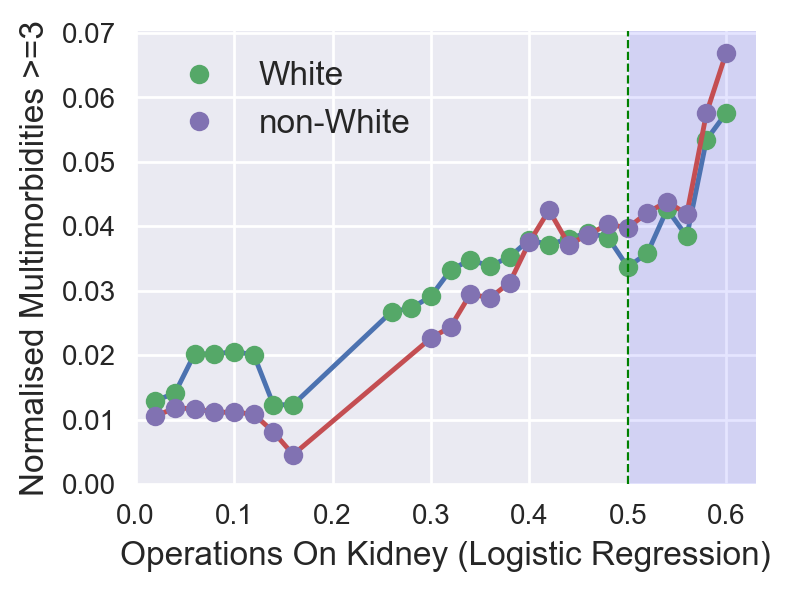} }%
\subfloat{\includegraphics[width=3.8cm]{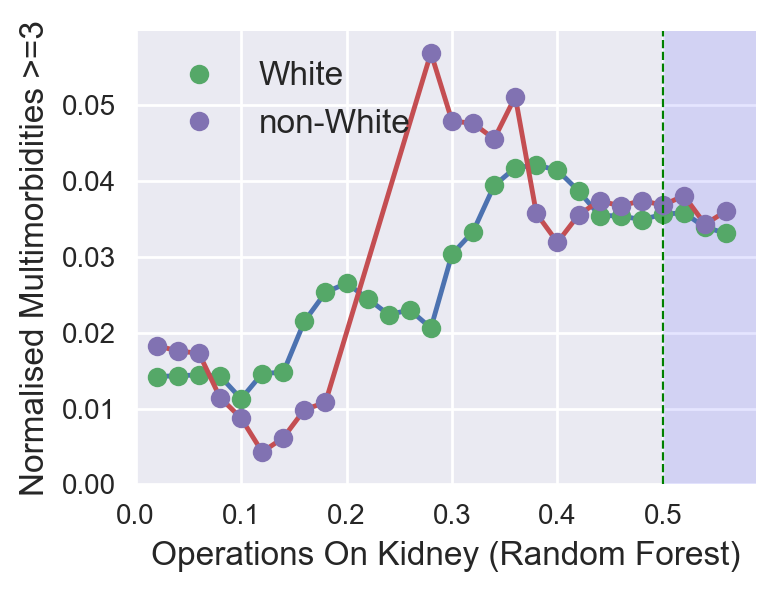} }%
\qquad
\subfloat{\includegraphics[width=3.8cm]{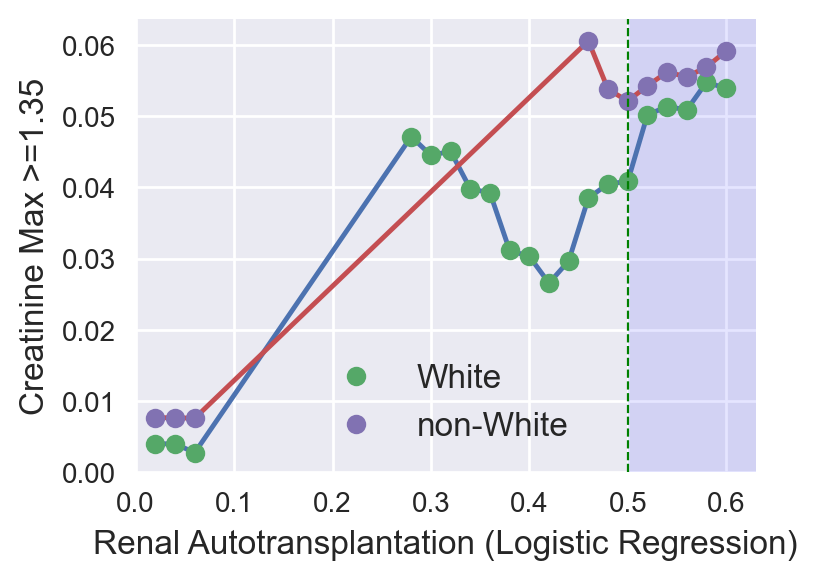} }%
\subfloat{\includegraphics[width=3.8cm]{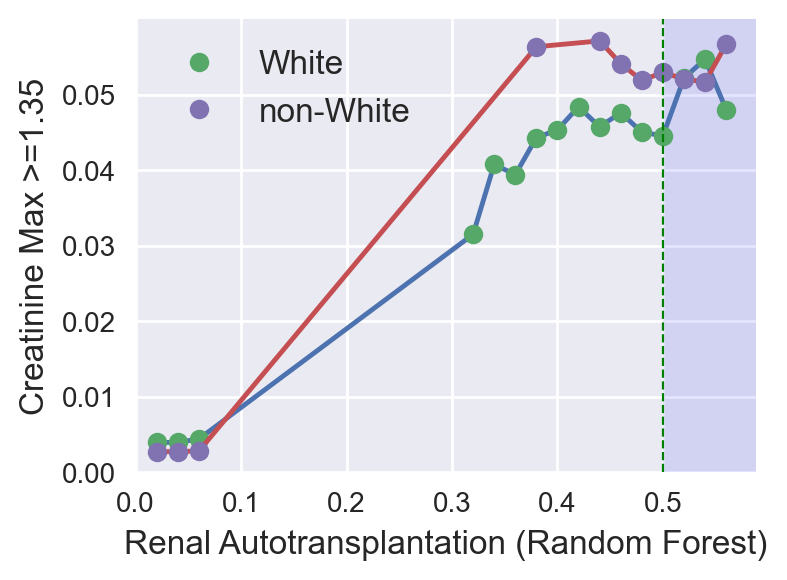} }%
\subfloat{\includegraphics[width=3.8cm]{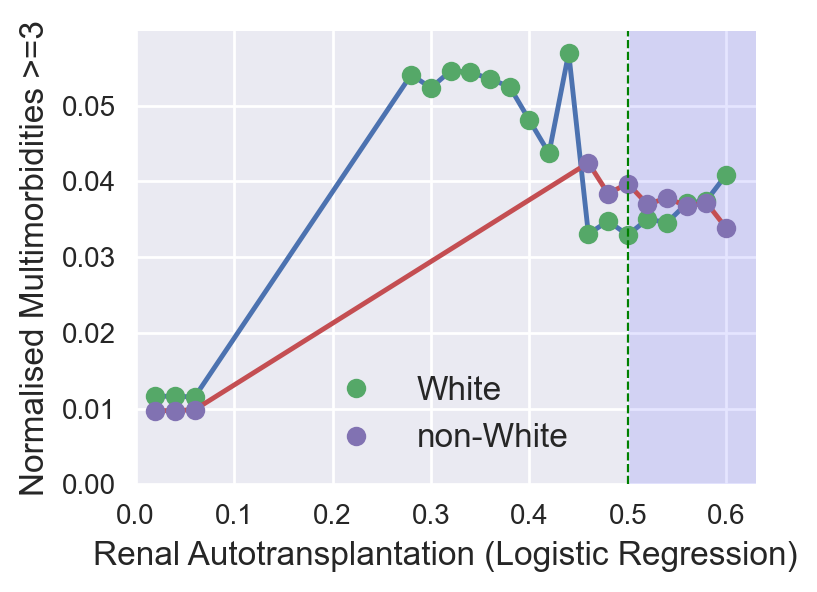} }%
\subfloat{\includegraphics[width=3.8cm]{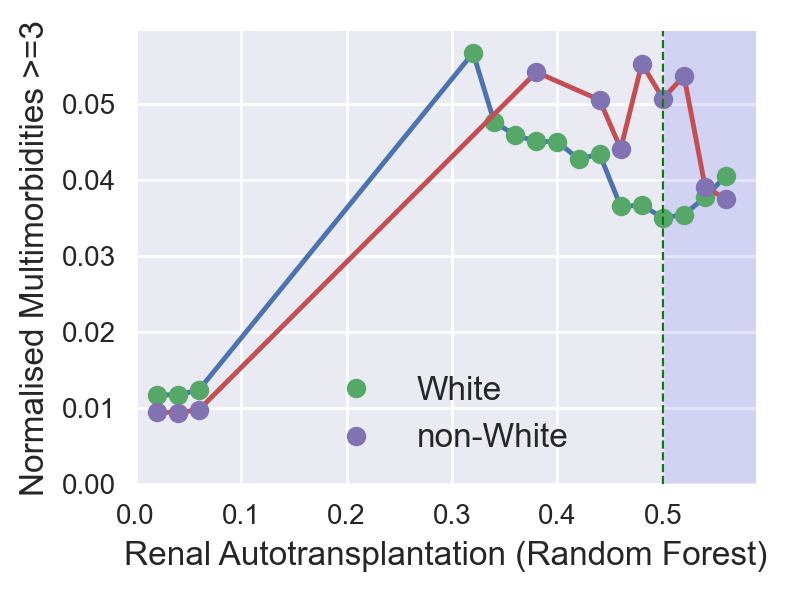} }%
\caption{Allocation-Deterioration Indices of four models trained for predicting the needs of kidney related surgeries. The top row is for a generic \emph{Operations on Kidney} and the bottom is for a particular \emph{Renal Autotransplantation}. The left two columns are those using \emph{deterioration index} defined on renal functions, while the right two are those using multimorbidities. In all cases, non-White patients are consistently more severe within the decision region (shaded area, allocation index $> 0.5$).}
\label{fig:modelineq}
\end{figure*}

Using inequalities quantified by Definition~\ref{def:aiineq}, Table~\ref{tab:modelieq1} summarises the inequality of \textbf{non-White vs White} induced by AI models for \emph{Kidney Operation}. Table~\ref{tab:modelieq} in the appendix shows full details of the two surgeries. Compared to inequality embedded in the database (of those who actually underwent surgeries), LR models exacerbated the inequality in 3 out of 4 assessments. RF tends to perform better in terms of mitigating the inequality in 3 out of 4 assessments, albeit very marginally in most cases. However, RF significantly exacerbated inequality more than 9 times on one occasion (see the last column of Appendix Table~\ref{tab:modelieq}). Overall, AI models induced inequalities in all cases and exacerbated inequalities severely in 3 out of 8 assessments. Figure~\ref{fig:modelineq1} illustrates one selected exemplar visualisation out of the eight total assessments (Appendix Figure~\ref{fig:modelineq} shows all eight). All these curves demonstrate a clear pattern that non-White patients are more deteriorated at the decision regions in all situations.

%Logistic regression models were trained to predict the procedures using three features: age, gender and the status of chronic kidney disease. The results showed that non-white groups have 9\% and 25\% more AUCs in AI model’s resource-allocation regions. These numbers quantify excess severity of the non-white group compared with whites when both groups are assigned the same risk scores by the models. In both cases, the models channelled the inequality embedded in the data (i.e., amongst patients who had gone through respective procedures). In the general kidney operation scenario, the model significantly exacerbated the inequality (from 1.58\% to 25.34\%). 

\section{Conclusion}
This paper proposes an Allocation-Deterioration framework that, to the best of our knowledge, is the first utility that visualises and quantifies health inequality induced by AI models and embedded in health datasets. Such a utility enables the evaluation, debugging and mitigation of inequality caused by AI technologies. While we focused on motivations and real-world data in the health domain, this framework is clearly generalisable and has much wider applications. An extensive set of experiments were conducted on two large, real-world datasets to assess its performances and reveal the existing (hidden) inequalities in different decision-making scenarios.

\clearpage
\bibliographystyle{named}
\bibliography{ijcai22}

\clearpage

\section*{Appendix of ``Quantifying Health Inequalities Induced by Data and AI Models"}

\section*{Ethics Statement}
Permission was granted by the data controllers to use the MIMIC-III and HiRID datasets. No personal data was processed in this study. 

% Algorithm~\ref{alg:adjb} gives the logic of doing left boundary adjustment, which also deals with pulse-like PDFs for discrete random variables.

% Algorithm~\ref{alg:curve} gives the approximation logic for generating Allocation-Deterioration Curves.

\RestyleAlgo{ruled}
\SetKwComment{Comment}{/* }{ */}

\begin{algorithm}[hbt!]
\caption{Left Boundary Adjustment}\label{alg:adjb}
\SetKwData{Left}{left}\SetKwData{This}{this}\SetKwData{Up}{up}
\SetKwFunction{Union}{Union}\SetKwFunction{FindCompress}{FindCompress}
\SetKwInOut{Input}{input}\SetKwInOut{Output}{output}

\Input {$E$: learned KDE\; $lb$: the lower bound\; 
$ub$: the upper bound\; 
$t$: value to adjust\; 
$t_p$: $\argmax(\{v|v \in M: v < t\})$ when $M$ is discrete and $t$ is not boundary, otherwise  $t$\; 
$\varepsilon$: a small constant like $1^{-10}$\; 
$V$: an empty array.} 
\Output {$\hat{t}$: the adjusted value for $t$}
\If{$len(V) = 0$}{
\Comment{get an evenly spaced numbers between $lb$ and $ub$ with a relatively big number $n$, e.g., $n=20 \times (ub - lb)$.}
$a \gets gen(lb, ub, n)$\;
$s \gets (ub - lb) / n$\;
\For{$i\leftarrow 1$ \KwTo len($a$)}{
$x_p \gets lb$\;
\If{$i > 1$}{
$x_p \gets a[i-1]$\;
}
$x \gets a[i]$\;
$p \gets exp(E(x))$\;
\While{$p \geq \varepsilon$ and $x > x_p$}{
$x \gets (x - s)$\;
$p \gets exp(E(x))$\;
}
\If{$exp(E(x)) < \varepsilon$}{
$V.add(x)$\;
}
}
}
$\hat{t} \gets \argmax(\{v|v \in V\colon v < t\})$\;
\If{$\hat{t} \leq t_p$}{
$\hat{t} \gets t$\;
}
return $\hat{t}$\; 
\end{algorithm}

\begin{algorithm}[]
\caption{Approximate A-D Curve}\label{alg:curve}
\SetKwInOut{Input}{input}\SetKwInOut{Output}{output}
\Input {$P$: the patient cohort\;
$d$: the deterioration index function\;
$m$: the measurement\;
$a$: the AI model as an allocator\;
$l$: a constant for smoothing\; 
$n$: a constant for specifying the number of points to be generated\;
$\nu$: a threshold for the minimal numbers of patients for deterioration estimation. 
} 
\Output {the curve}
$C \gets []$\;
$X \gets gen(0, 1, n)$\;
\For{$x$ in $X$}{
$\hat{P} \gets \{p|p\in P: (x-l) \leq a(p) < (x+l)\}$\;
\If{$|\hat{P}|\geq \nu$}{
$C.add((x, d(\hat{P};m)))$\;
}
}
return $C$
\end{algorithm}

% \section{Cohort Characteristics}
% Table~\ref{tab:ptch1} and \ref{tab:ptch2} give the characteristics of the two patielt cohorts from the MIMIC III dataset.

% \section{Result related contents}
% Table~\ref{tab:10synth} illustrates the experiment results on 10 synthetic datasets for evaluting the framework's ability on detecting NO inequalities. The T test results (p-values) show in all three measurements our framework detects non-substantial inequality.

% Figure~\ref{fig:increaseEval} shows the results of controlled inequality improvements. It demonstrates that our model could correctly capture the levels of inequality in the data.

% Figure~\ref{fig:modelineq} depicts the Allocation-Deterioration Indices of four models trained for predicting the needs of kidney related surgeries. 

\begin{table}[]
\center
\begin{tabular}{ccl}
\multicolumn{3}{c}{Health inequality assessments on synthetic datasets}                                                                                      \\ \hline
\multicolumn{1}{l}{Measurement}                                                & mean [95\% CI]                                  & $p$-value \\ \hline \hline
\multicolumn{1}{c|}{\begin{tabular}[c]{@{}c@{}}Creatinine \\ max\end{tabular}} & \multicolumn{1}{c|}{0.044 {[}-0.083,  0.130{]}} & 0.0664  \\ \hline
\multicolumn{1}{c|}{\begin{tabular}[c]{@{}c@{}}Creatinine \\ min\end{tabular}} & \multicolumn{1}{c|}{0.024 {[}-0.266,  0.302{]}} & 0.7084  \\ \hline
\multicolumn{1}{c|}{\begin{tabular}[c]{@{}c@{}}ALT \\ max\end{tabular}}        & \multicolumn{1}{c|}{0.033 {[}-0.157,  0.182{]}} & 0.4231  \\ \hline
\end{tabular}
\caption{Overall inequality of \textbf{female vs male} quantified on 10 synthetic datasets, where there should be no inequality overall.}
\label{tab:10synth}
\end{table}

\begin{figure*}[]
\caption{Inequality Quantification Evaluation on synthetic data: y-axis is the inequality quantity of female vs male. x-axis is the percentage of controlled improvements on readings of the female subcohort.  Y-value of each point is the mean value of 10 runs on the same x-value, i.e., \% of improvement. Shaded areas denote 25-75\% quantile regions.}
\label{fig:increaseEval}
\begin{center}
\includegraphics[width=12cm]{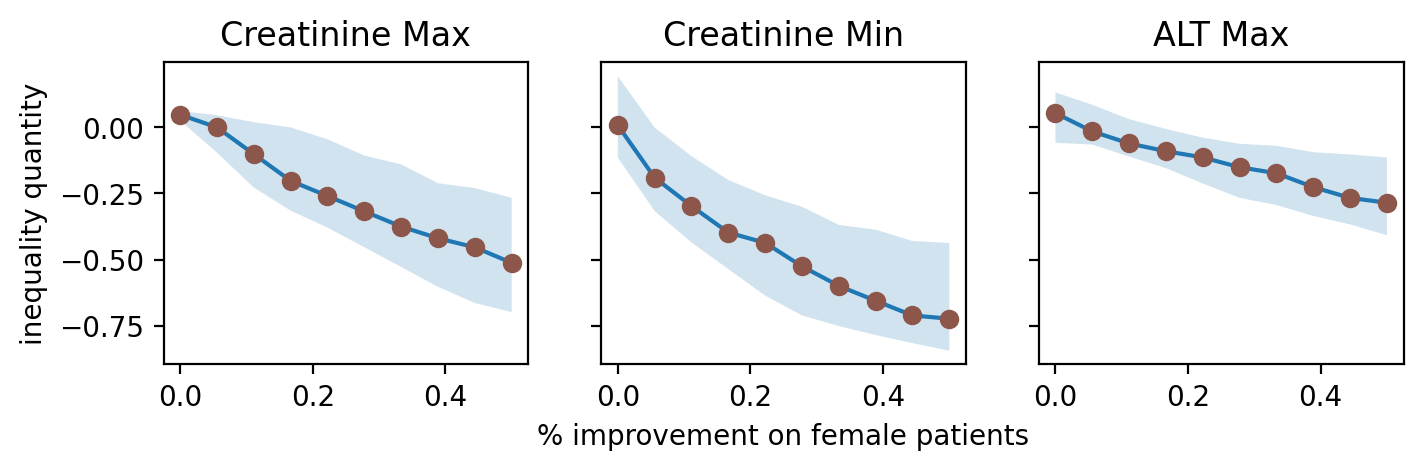}
\end{center}
\end{figure*}

\begin{figure*}[]
\subfloat{\includegraphics[width=7.71cm]{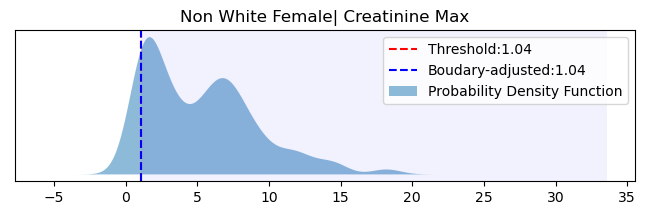} }%
% \qquad
\subfloat{\includegraphics[width=7.71cm]{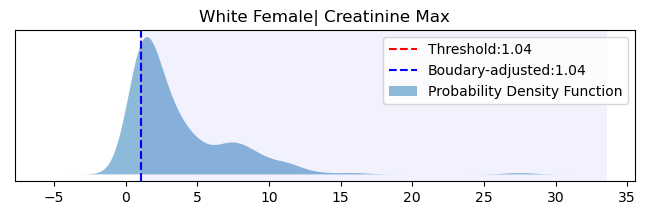} }%
\qquad
\subfloat{\includegraphics[width=7.71cm]{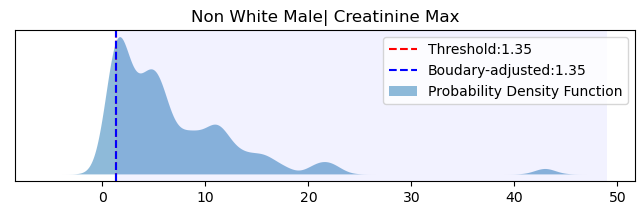} }%
% \qquad
\subfloat{\includegraphics[width=7.71cm]{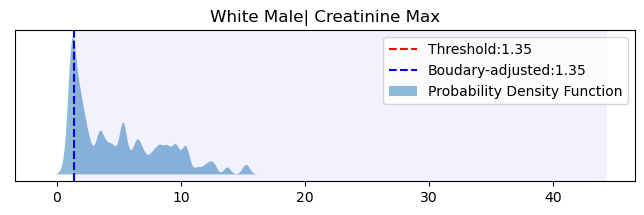} }%
\caption{Probability density functions for quantifying inequalities of \textbf{non-White vs White} in the scenario of kidney operations in MIMIC-III dataset. Dashed lines denote thresholds (i.e., boundary values of abnormal readings) for computing deterioration index. Shaded area are regions where the probability integral happens for getting the deterioration index. The above two figures are females, which illustrate an inequality of 35.06\%. The bottom two are males, where there is an inequality of 19.94\%.}
\label{fig:mimiccreatinine}
\end{figure*}

\begin{table*}[]
\begin{tabular}{rcrcr||crcr}
\cline{2-9}
\multicolumn{1}{l}{}                                                     & \multicolumn{4}{c||}{Kidney operation}                                                                                          & \multicolumn{4}{c}{Renal Autotransplantation}                                                               \\ \cline{2-9} 
\multicolumn{1}{l}{}                                                     & \multicolumn{2}{l}{Creatinine Max}                            & \multicolumn{2}{l||}{Normalised MM}                             & \multicolumn{2}{l}{Creatinine Max}                  & \multicolumn{2}{l}{Normalised MM}                     \\ \cline{2-9} 
DB inequality                                                            & \multicolumn{2}{c}{29.10\%}                                   & \multicolumn{2}{c||}{7.62\%}                                    & \multicolumn{2}{c}{16.08\%}                         & \multicolumn{2}{c}{2.58\%}                            \\ \hline \hline
Models                                                                   & LR                                   & \multicolumn{1}{c}{RF} & LR                                   & \multicolumn{1}{c||}{RF} & LR                         & \multicolumn{1}{c}{RF} & LR                           & \multicolumn{1}{c}{RF} \\ \hline
\begin{tabular}[c]{@{}r@{}}Inequality at \\ Decision Region\end{tabular} & \multicolumn{1}{r}{\textbf{37.58\%}} & 22.15\%                & \multicolumn{1}{r}{\textbf{10.52\%}} & 4.54\%                  & \multicolumn{1}{r}{9.13\%} & 3.51\%                 & \multicolumn{1}{r}{2.45\%}   & \textbf{23.36\%}       \\ \hline
\begin{tabular}[c]{@{}r@{}}Inequality at \\ the whole area\end{tabular}  & \multicolumn{1}{r}{16.17\%}           & 30.21\%                & \multicolumn{1}{r}{-11.8\%}          & 9.65\%                  & \multicolumn{1}{r}{14.73\%} & 22.70\%                & \multicolumn{1}{r}{-26.10\%} & 0.20\%                 \\ \hline
\end{tabular}
\caption{Inequality of \textbf{non-White vs White patients} channelled and exacerbated by AI models in two decision-making scenarios of kidney related operations in the MIMIC-III dataset. \emph{DB inequality} row gives the DB embedded inequality quantities of relevant measurements. \emph{Inequality at Decision Region} is the area between A-D curves within the region where a model suggesting surgery, while \emph{Inequality at the whole area} is the area between two curves overall.}
\label{tab:modelieq}
\end{table*}

% \section{Experiment Setup}
% Table~\ref{tab:hyperp} gives the reproducible hyperparameters and other related setups for training AI models for predicting the needs of kidney operations in the MIMIC III dataset.

\begin{table*}[]
\center
\begin{tabular}{l|l}
\hline
\textbf{Attributes}                                                            & \textbf{Details}                                                                                                                                                                                                                                                                          \\ \hline
Feature List                                                                   & \begin{tabular}[c]{@{}l@{}}{[}'age', 'Chronic kidney disease', 'gender',\\ 'Leukemia', 'cirrhosis', 'Infection'{]}\end{tabular}                                                                                                                                                           \\ \hline
\begin{tabular}[c]{@{}l@{}}Random Forest \\ Hyper-parameters\end{tabular}      & \begin{tabular}[c]{@{}l@{}}tuned\_parameters = \{\\             'n\_estimators': {[}50, 100, 200{]},\\             'max\_depth': {[}5, 10, 20, 50{]}\\     \}\end{tabular}                                                                                                                \\ \hline
\begin{tabular}[c]{@{}l@{}}Logistic Regression\\ Hyper-parameters\end{tabular} & \begin{tabular}[c]{@{}l@{}}tuned\_parameters = \{\\             'penalty': {[}'l1', 'l2'{]},\\             'C': {[} \#.001, .01, \\                   .1, 1, 10, 100, 1000{]},\\             'max\_iter': {[}100, 150{]},\\             'solver': {[}'liblinear'{]}\\     \}\end{tabular} \\ \hline
Random state                                                                   & 1                                                                                                                                                                                                                                                                                         \\ \hline
\end{tabular}
\caption{AI Model's hyperparameters and other reproducible setups}
\label{tab:hyperp}
\end{table*}

\begin{table*}[]
\center
\begin{tabular}{lll}
\hline
\multicolumn{3}{c}{\textbf{Renal Autotransplantation}}                                                                   \\ \hline
\multicolumn{1}{l|}{}                           & \multicolumn{1}{l|}{Case}                    & Control                 \\ \hline
\multicolumn{1}{l|}{N}                          & \multicolumn{1}{l|}{146}                     & 438                     \\ \hline
\multicolumn{1}{l|}{Gender(male)}               & \multicolumn{1}{l|}{83 (56.8\%)}             & 286 (65.3\%)            \\ \hline
\multicolumn{1}{l|}{Age}                        & \multicolumn{1}{l|}{53.31 {[}47.00-60.75{]}} & 53.47 {[}47.00-61.00{]} \\ \hline
\multicolumn{1}{l|}{Clinical attributes}        & \multicolumn{1}{l|}{}                        &                         \\ \hline
\multicolumn{1}{l|}{Length of Stay(days)}       & \multicolumn{1}{l|}{10.88 {[}6.00-14.00{]}}  & 8.07 {[}3.00-11.00{]}   \\ \hline
\multicolumn{1}{l|}{Death}                      & \multicolumn{1}{l|}{5 (3.4\%)}               & 29 (6.6\%)              \\ \hline
\multicolumn{1}{l|}{CKD}                        & \multicolumn{1}{l|}{145 (99.3\%)}            & 157 (35.8\%)            \\ \hline
\multicolumn{1}{l|}{Cirrhosis}                  & \multicolumn{1}{l|}{25 (17.1\%)}             & 35 (8.0\%)              \\ \hline
\multicolumn{1}{l|}{Infection}                  & \multicolumn{1}{l|}{37 (25.3\%)}             & 90 (20.5\%)             \\ \hline
\multicolumn{1}{l|}{Number of multimorbidities} & \multicolumn{1}{l|}{4.27 {[}3.00-5.00{]}}    & 2.83 {[}1.00-4.00{]}    \\ \hline
\end{tabular}
\caption{Characteristics of the study cohorts for the Renal Autotransplantation prediction task. The case cohort is identified from the MIMIC III database using ICD-9 code 55.61 and the control cohort is matched using similar age (+/- 3) with 1:3 ratio. }
\label{tab:ptch1}
\end{table*}

\begin{table*}
\center
\begin{tabular}{lll}
\hline
\multicolumn{3}{c}{\textbf{Operations on Kidney}}                                                                        \\ \hline
\multicolumn{1}{l|}{}                           & \multicolumn{1}{l|}{Case}                    & Control                 \\ \hline
\multicolumn{1}{l|}{N}                          & \multicolumn{1}{l|}{584}                     & 1752                    \\ \hline
\multicolumn{1}{l|}{Gender(male)}               & \multicolumn{1}{l|}{293 (50.2\%)}            & 1,018 (58.1\%)          \\ \hline
\multicolumn{1}{l|}{Age}                        & \multicolumn{1}{l|}{58.78 {[}49.00-69.00{]}} & 58.91 {[}49.00-70.00{]} \\ \hline
\multicolumn{1}{l|}{Clinical attributes}        & \multicolumn{1}{l|}{}                        &                         \\ \hline
\multicolumn{1}{l|}{Length of Stay(days)}       & \multicolumn{1}{l|}{10.43 {[}5.00-14.00{]}}  & 8.24 {[}3.00-11.00{]}   \\ \hline
\multicolumn{1}{l|}{Death}                      & \multicolumn{1}{l|}{34 (5.8\%)}              & 165 (9.4\%)             \\ \hline
\multicolumn{1}{l|}{CKD}                        & \multicolumn{1}{l|}{537 (92.0\%)}            & 665 (38.0\%)            \\ \hline
\multicolumn{1}{l|}{Cirrhosis}                  & \multicolumn{1}{l|}{35 (6.0\%)}              & 117 (6.7\%)             \\ \hline
\multicolumn{1}{l|}{Infection}                  & \multicolumn{1}{l|}{219 (37.5\%)}            & 366 (20.9\%)            \\ \hline
\multicolumn{1}{l|}{Number of multimorbidities} & \multicolumn{1}{l|}{3.74 {[}2.00-5.00{]}}    & 3.18 {[}1.00-5.00{]}    \\ \hline
\end{tabular}
\caption{Baseline Characteristics of the study cohorts for the Operations on Kidney prediction task. The case cohort is identified from the MIMIC III database using ICD-9 codes of 55.xx and the control cohort is matched using similar age (+/- 3) with 1:3 ratio. }
\label{tab:ptch2}
\end{table*}

\end{document}